\documentclass[twocolumn,%showpacs,
prl
,superscriptaddress
]{revtex4-1}

\usepackage{graphicx}
\usepackage{amssymb}
\usepackage{amsmath}
\usepackage{color}
\usepackage{comment}

\begin{document}

\title{Bias, Belief and Consensus: Collective opinion formation on fluctuating networks}

\author{Vudtiwat Ngampruetikorn}
\affiliation{Biological Physics Theory Unit, Okinawa Institute of Science and Technology Graduate University, Onna, Okinawa 904 0495, Japan} %

\author{Greg J. Stephens}
\affiliation{Biological Physics Theory Unit, Okinawa Institute of Science and Technology Graduate University, Onna, Okinawa 904 0495, Japan} %
\affiliation{Department of Physics \& Astronomy, Vrije Universiteit Amsterdam, 1081HV Amsterdam, The Netherlands}

\date{\today}

\begin{abstract}

With the advent of online networks, societies are substantially more connected with individual members able to easily modify and maintain their own social links.  Here, we show that active network maintenance exposes agents to confirmation bias, the tendency to confirm one's beliefs, and we explore how this affects collective opinion formation. We introduce a model of binary opinion dynamics on a complex network with fast, stochastic rewiring and show that confirmation bias induces a segregation of individuals with different opinions.  We use the dynamics of global opinion to generally categorize opinion update rules and find that confirmation bias always stabilizes the consensus state. Finally, we show that the time to reach consensus has a non-monotonic dependence on the magnitude of the bias, suggesting a novel avenue for large-scale opinion engineering.
\end{abstract}

\maketitle

A society is molded by the opinions of its members and with enough time pervasive opinions become the basis of customs and culture. 
However even the most deeply-rooted ideas can be challenged and occasionally, even rapidly, dethroned such as occurred recently with the attitudes towards gay marriage in the US \cite{gaymarriage}. 
On the other hand, fragmented, polarized opinions can and do persist so that whether and how a belief emerges as a social norm is a fundamental and important question~\cite{Acemoglu:2011}. 

While consensus is a collective state, the existence and dynamics of this state depends on how individuals receive, process and respond to information, all of which are influenced by cognitive bias, i.e.,  systematic errors in subjective judgements~\cite{Tversky:1974}. 
Among various cognitive biases, `confirmation bias' -- the tendency to confirm one's beliefs \cite{Nickerson:1998} -- is not only particularly relevant to opinion formation but also sensitive to new communication technology~\cite{Rosenblat:2004,Stroud:2008,Valentino:2009,Garrett:2009,Bakshy:2015}. In online social networks, for example, it only takes seconds to silence disagreeing views or to seek out new like-minded contacts. Indeed, the increasing ability to fashion and maintain our own social circles regardless of geographic proximity amplifies the effects of a general preference for agreeing information sources, a confirmation bias which is known as `selective exposure' \cite{Frey:1986,Stroud:2008,Valentino:2009,Garrett:2009}.  Individuals are additionally subject to `biased interpretation' \cite{Lord:1979,Griffin:1992}, the inclination to discount information undermining one's views. Understanding how these and other cognitive biases affect collective opinion formation is likely to be increasingly important to understanding social discourse in modern societies.

Here, we explore how confirmation bias can effect the formation, stability and dynamics of the consensus state.  
For discrete opinions, the quantitative analysis of consensus dynamics can be traced to various kinetic Ising models~\cite{Galam:1982,Galam:1991,Deffuant:2000,Krapivsky:2003,Mobilia:2003,Sood:2005,Suchecki:2005,Castellano:2009}, including the pioneering work of the voter model \cite{Clifford:1973,Holley:1975} which is exactly solvable in all dimensions. 
The hallmarks of the voter model include (see, e.g., Ref.~\cite{Krapivsky:2010}): inevitability of consensus; probabilistic conservation of average opinion; and how dimensionality $d$ qualitatively affects the time to consensus: $T_C\sim N^2$ for $d=1$, $N\ln N$ for $d=2$ and $N$ for $d>2$ with $N$ denoting the system size.  
Latter analyses have considered social-network-inspired refinements, including complex, more natural network topologies~\cite{Castellano:2003,Vilone:2004,Sood:2005,Suchecki:2005,Sood:2008} (see, e.g., Ref.~\cite{Castellano:2009} for a review) and, more recently, the beginning of a fully dynamical approach~\cite{Holme:2006,Benczik:2009,Zschaler:2012,Durrett:2012,Rogers:2013}.

In distinction to imposing and refining static network structures, we endow the links with dynamics so that collective opinions and the changing social network in which they sit both emerge and coevolve.  
Agents update their opinion based on their neighbors and links form and disappear with differing dynamics depending on whether they connect agreeing or disagreeing agents. We find that confirmation bias in the form of asymmetric link dynamics leads naturally to opinion segregation. In the limit of large bias, we show that opinion polarization is unstable and consensus is stable and this result is universal within a general family of update rules.  We further show that although strong bias in general increases consensus time, weaker bias could speed up consensus formation, a surprising result with implications for active network engineering.  Finally we discuss the implications of our results for various social systems.

%%%%%%%%%
\begin{figure}
\centering
\includegraphics[width=\linewidth]{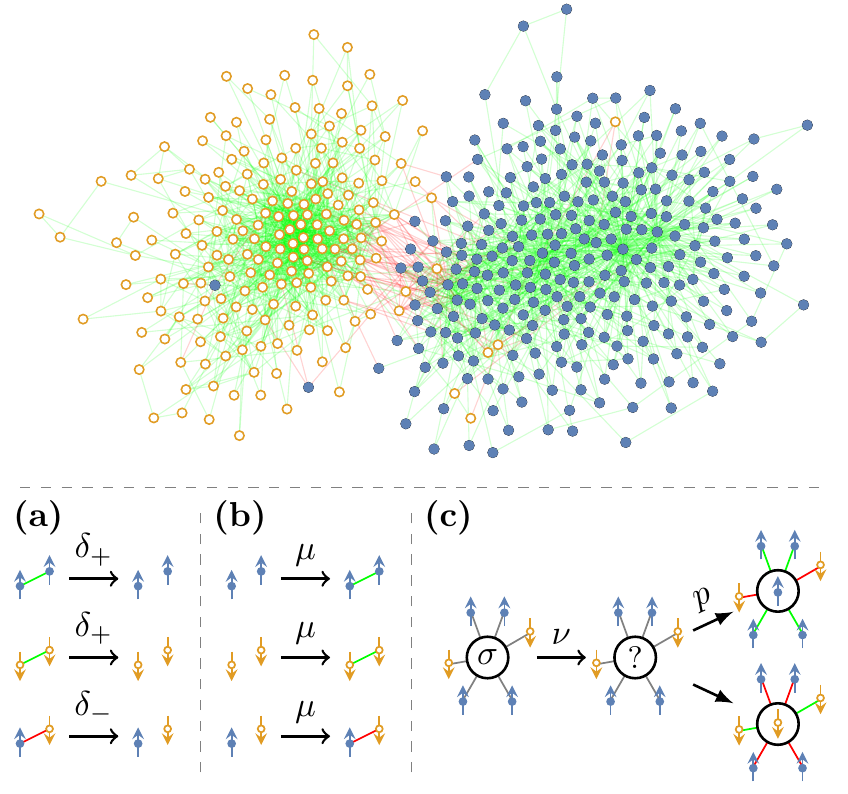}
\caption{
We model confirmation bias in collective opinion formation as heterogenous link dynamics coupled with opinion updates on a complex network. 
{\bf Top} -- {\it Example Opinion Network:}
Segregation is evident in a network of political weblogs (data from \cite{Adamic:2005}), labelled according to their inclinations: liberal (empty circles) or conservative (filled circles).\footnote{To represent only the most active blogs, we remove nodes of degree one until the minimum degree is two.}
{\bf Bottom} -- {\it Model:}
\textbf{(a)}  
All links have a finite lifetime. 
Agreeing links (green) connect like-minded agents and decay at a rate $\delta_+$ while disagreeing links (red) decay at a rate $\delta_-> \delta_+$. 
\textbf{(b)} 
New links are formed at a rate $\mu$.
\textbf{(c)} 
`Spineless' agents discard their current opinions at a rate $\nu$ and immediately adopt $\uparrow$ opinion with a probability $p$, determined purely by the opinions of their nearest neighbors.  Opinion updating and rewiring are intricately coupled; the local network structure, governed by rewiring dynamics, dictates opinion updating and opinion configurations determines the decay rates in rewiring dynamics.
\label{fig:model}
}
\end{figure}
%%%%%%%%

We introduce a model of opinion dynamics to encompass both bias and opinion changes [Fig.~\ref{fig:model}]. 
Here, as in previous approaches \cite{Sood:2005,Suchecki:2005,Durrett:2012,Rogers:2013}, agents are represented as nodes and carry a binary opinion $\uparrow$ ($\sigma=1$) or $\downarrow$ ($\sigma=-1$). 
The network, however, isn't static~\cite{Holme:2006,Benczik:2009,Zschaler:2012,Durrett:2012,Rogers:2013}. 
Instead, links disappear randomly so that connections between agents with the same opinion (agreeing links) decay at a rate $\delta_+$ while those between opposite opinions (disagreeing links) decay at a faster rate $\delta_->\delta_+$ [Fig.~\ref{fig:model}(a)]. 
This difference in rates encodes the social forces that push individuals towards agreement and away from cognitive conflict \cite{Lazarsfeld:1954,McPherson:2001}. 
As disagreeing links are shorter lived, islands of agreement form spontaneously around each agent and these islands are the manifestation of confirmation bias. 
Finally, we allow for spontaneous new connections between any two unconnected nodes and this occurs with rate $\mu$ [Fig.~\ref{fig:model}(b)] which we choose proportional to $1/N$ to maintain a constant mean degree across different network sizes.

All agents reevaluate their current opinions at a rate $\nu$ after which they adopt an $\uparrow$ opinion with probability $p$ [Fig.~\ref{fig:model}(c)]. 
We consider only `spineless' agents -- those whose opinion is determined purely by the opinions of their neighbors. 
We will return to more complex decision rules such as Bayesian updating (see, e.g., Refs.~\cite{DeMarzo:2003,Acemoglu:2011}) in the discussion. 
In distinction to previous work on opinion networks (though see \cite{Oliveira:1993,Drouffe:1999,Mobilia:2003a} on Ising systems), we characterize the update probability $p$ through a generalized family of functions so that commonly analyzed update processes such as `majority rule' and `proportional rule ' appear as limiting cases. 
We note that the sculpting of the network through the dynamics of confirmation bias detailed above means that the composition of neighborhoods at the time of opinion reevaluation depends on the previous opinion; agents deposit their memory in the network structure and are effectively less spineless than those in a static network.

Real social networks are becoming larger and more connected as links are easy to create and maintain.  Facebook, for example, currently consists of over a billion users, each with an average of hundreds of friends \cite{Ugander:2011}. 
Thus we consider our model in the limits of large population and average degree $1\ll \bar k \ll N$ where a mean-field approximation is accurate. Online networks also greatly facilitate link maintenance so that it is relatively easy to acquire new neighbors and silence existing ones.  
Here we assume that network rewiring is much faster than and therefore decoupled from opinion updating: between two opinion updates, there is always enough time for the network to reach `equilibrium wiring'.

%%%%%%%%%
\begin{figure}
\centering
\includegraphics[width=\linewidth]{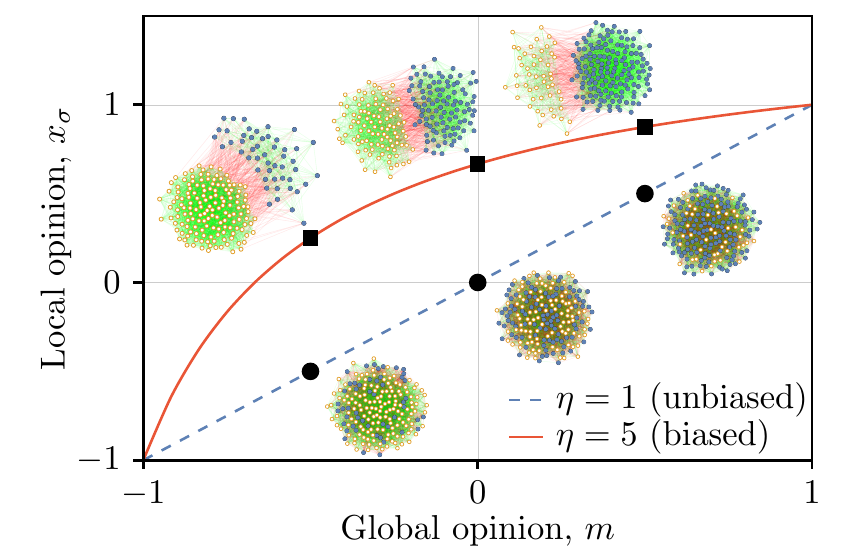}
\caption{Confirmation bias sculpts local social circles and induces segregation of opinions. As seen by a reference agent with $\uparrow$ opinion, we show local opinion as a function of global opinion for biased ($\eta>1$, solid) and unbiased ($\eta=1$, dashed) network dynamics. Snapshots depict exemplar networks and illustrate that unbiased systems (disks)
are well-mixed while biased systems (squares) are segregated. The local opinion $x_\sigma$ is the average opinion of neighbors [Eq.~\eqref{eq:x_s}] and the global opinion $m$ is the average opinion of all agents [Eq.~\eqref{eq:definition}]. 
\label{fig:rewiring}
}
\end{figure}
%%%%%%%%%

The wiring is encoded in the adjacency matrix $A$ so that $A_{ij}=1$ if there exists a link between agents $i$ and $j$ and $A_{ij}=0$ otherwise. 
Given a static opinion configuration, the master equation for $A$ is given by
\begin{equation}
\dot P(A_{ij}=1) = -\delta_{ij} P(A_{ij}=1) + \mu P(A_{ij}=0),
\end{equation}
where $\delta_{ij} = \delta_\pm$ for $\sigma_i \sigma_j = \pm1$. 
This yields a solution 
\begin{equation}
P(A_{ij}=1) \approx {\mu}/{\delta_{ij}} + C e^{-\delta_{ij}t},
\end{equation}
where we have taken the limit $N\to\infty$ and $C$ is the constant of integration. 
The transient dynamics defines the adiabatic approximation so that opinion updating is slow if $\min \delta_\pm \gg \nu$.
In equilibrium, the probability that a link between agreeing and disagreeing agents exists is $P_{+} = {\mu}/{\delta_+}$ and $P_{-} = {\mu}/{\delta_-}$, respectively.
Note that although the equilibrium wiring does not contain explicit time dependence, it is determined by the opinion configuration which does evolve (slowly) in time.

The mean numbers of agreeing and disagreeing links around a $\sigma$-agent (an agent with opinion $\sigma$) are given by  $(N_\sigma-1)P_+$ and  $N_{-\sigma}P_-$, respectively. In the mean field approximation, the average neighbor opinion around this agent is given by 
\begin{equation}
x_\sigma \approx \sigma
\frac{ N_\sigma P_+ -  N_{-\sigma}P_-}{ N_\sigma P_+ +  N_{-\sigma}P_-}
=
\sigma \frac{ (\sigma+m)\eta -   (\sigma-m)}{(\sigma+m)\eta +   (\sigma-m)}
\label{eq:x_s}
\end{equation}
where the global opinion $m$ and the parameterisation of confirmation bias $\eta$ are defined as follows 
\begin{equation}
m\equiv\frac{N_\uparrow-N_\downarrow}{N_\uparrow+N_\downarrow}
\quad\text{and}\quad
\eta \equiv \frac{P_+}{P_-}.
\label{eq:definition}
\end{equation}
The system is biased, exhibiting confirmation bias, if $\eta>1$ and unbiased if $\eta=1$. 
Our model also allows for anti-confirmation bias ($\eta<1$) but no strong evidence exists for such an anti-social behavior on large networks and we omit this case in the following analysis.

In Figure~\ref{fig:rewiring} we show how asymmetric link dynamics (which underlie confirmation bias in our model) affects network structure and local opinion. 
When $\eta = 1$, the system is well-mixed and the local and global opinions are equal, $x_\sigma=m$. 
For $\eta>1$, on the other hand, agents of differing opinions segregate as each forms a more agreeable neighborhood, $x_\downarrow\le m\le x_\uparrow$.

%%%%%%%%
\begin{figure}
\centering
\includegraphics[width=\linewidth]{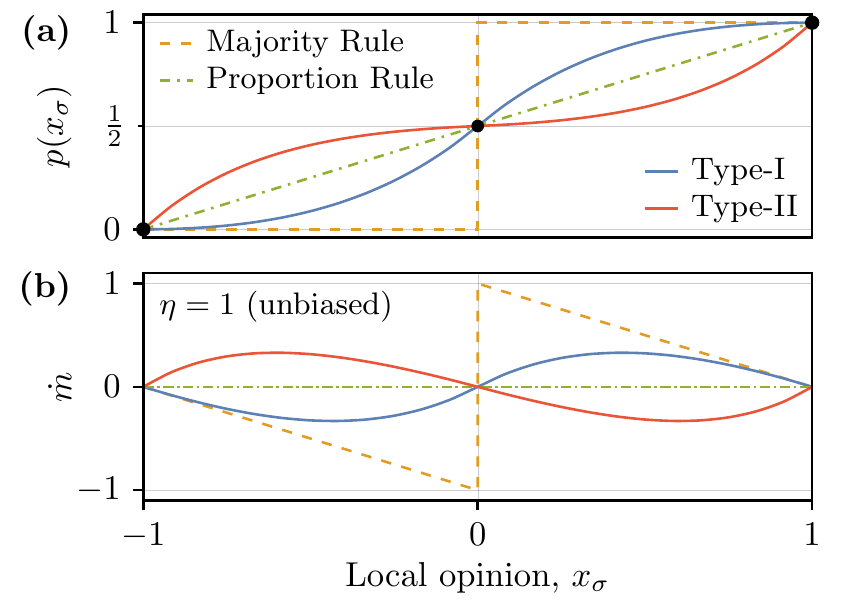}
\caption{Opinion update rules can be generally classified based on global opinion dynamics in unbiased systems. 
\textbf{(a)}
The probability of adopting $\uparrow$ opinion [see Fig.~\ref{fig:model}(c)]. 
Given the constraints of spinelessness (solid disks), conformity ($p'(x)\ge 0$) and a mild assumption of a single inflection point, \textit{all} update rules can be classified as type-I (blue) or type-II (red), based on the resulting dynamics (see panel below). 
Also shown are the well-known special cases of our classification. 
The majority rule (dashed) is the extreme limit of type-I rules 
while the proportion rule (dash dotted) -- also known as the voter model -- 
is at the interface between the two types of rules. 
\textbf{(b)}
Dynamical phase space in unbiased systems for the same update rules as in \textbf{(a)}.
Consensus is stable for type-I rules and unstable for type-II rules.
For the proportion rule, the system is completely static and fluctuations are important in finite systems. 
\label{fig:rules}
}
\end{figure}
%%%%%%%%

Spineless agents possess neither intrinsic beliefs nor memories so that their update rules depend entirely on local opinions; 
the probability of adopting $\uparrow$ opinion is given by $p_\sigma = p(x_\sigma)$ for an agent whose previous opinion is $\sigma$. 
Spinelessness also implies invariance under opinion exchange, $\uparrow\leftrightarrow\downarrow$, thus $p(-x)=1-p(x)$. 
Once there is consensus among neighbors, spineless agents always adopt the consensus opinion hence $p(-1)=0$ and $p(1)=1$. 
In Fig.~\ref{fig:rules}(a) we illustrate a variety of update rules, including the common examples of majority and  proportional rule and we show later that these rules can be simply classified based on the overall network dynamics. 
In all cases, we further assume that agents are conforming so that they adopt an opinion with a probability that increases with the local fraction of that opinion, $p'(x)\ge 0$. 

Collective opinion states emerge as fixed points of the global dynamics, 
\begin{align}
\dot m = -(1+m)(1-p(x_\uparrow)) + (1-m)p(x_\downarrow).
\label{eq:dynamics}
\end{align}
Here $m$ denotes the global opinion and $x_{\sigma=\{\uparrow,\downarrow\}}$ the local opinion around $\sigma$-agents.  We have also set $\nu\equiv 1$ so that all other timescales are expressed in these units. 
The first term arises from $\uparrow$-to-$\downarrow$ conversions and the other describes $\downarrow$-to-$\uparrow$ conversions. 
We note that the dynamics is invariant under opinion exchange, $p(x_\sigma)\to1-p(x_{-\sigma})$ and $m\to-m$, as required by spinelessness. 
Two fixed points are always present. 
They correspond to consensus, where $m^*=\pm1$ and $x^*_\sigma=\pm1$, and the polarized state, where $m^*=0$ and, owing to spinelessness, $x^*_\uparrow=-x^*_\downarrow$. 
The asterisks denote local and global opinions at the corresponding fixed points.
In general, the behavior of the dynamics is determined by the nature (e.g., attractive or repulsive) of these fixed points. 

The stability of each fixed point depends on both the opinion update rules and the severity of confirmation bias. 
To separate the role of the update rules, we first focus on the unbiased cases ($\eta=1$). 
Here we find, under the mild assumption of a single inflection point, that \textit{all} update rules can be classified as type-I or type-II [Fig.~\ref{fig:rules}(a)], depending on the resulting dynamics [Fig.~\ref{fig:rules}(b)].
Concave when $x>0$, type-I rules drive the systems to consensus, $m=\pm1$. 
Convex when $x>0$, type-II rules favor the polarised state, $m=0$. 
Neither concave or convex, the interface between the two classes of rules corresponds to the proportion rule, also known as the voter model \cite{Clifford:1973,Holley:1975}. 
This special case leads to completely static systems in our mean field approximation. 
The majority rule or zero-temperature Glauber kinetics~\cite{Glauber:1963}, in which agents always join the majority in their neighborhoods, is an extreme limit of type-I. 

For the biased dynamics ($\eta>1$), we find that strong confirmation bias stabilizes consensus and destabilizes opinion polarization and this result is true {\it regardless of the update rules.} 
To see how bias stabilizes consensus, we consider a one-agent perturbation from consensus. 
The minority agent is always surrounded by majority agents and the minority-to-majority conversion rate is thus fixed, regardless of bias.
As biased majority agents discount the presence of the minority, the majority-to-minority conversion rate approaches zero; hence, consensus is stable. 
More precisely, near the consensus fixed point, we have
\begin{align}
\left.\frac{d\dot m}{dm}\right|_{m=1} = -1+2p'(1)f'(1),
\label{eq:dmdotdm1}
\end{align}
where $x_\sigma\equiv \sigma f(\sigma m)$ describes the relation between local and global opinions. 
In general, confirmation bias implies $f'(1)\le1$ (see also Fig.~\ref{fig:rewiring}) and, with strong bias, small minority fraction will not change local opinions around majority-agents, i.e., $f'(1)\to0$ as $\eta\to\infty$. 
Consequently $d\dot m/dm|_{m=1} \rightarrow-1$ from above (since we always have $p'(1)\ge0$), i.e.,  consensus stability increases with bias. 

To understand the dynamics near the polarized state, we recall that in general $f(m)$ is given by Eq.~\eqref{eq:x_s}. 
In the limit $\eta\gg1$, this yields 
\begin{align}
\left.\frac{d\dot m}{dm}\right|_{m=0} 
\,\overset{\eta\gg1}{\approx}\,
p'(1)\frac{4}{\eta}  + \mathcal{O}\left(\frac{1}{\eta^2}\right).
\label{eq:dmdotdm0}
\end{align}
That is, opinion polarization is always unstable in this limit since $p'(1)\ge0$. 
We see that, near both fixed points, strong bias leads to \textit{universal} dynamics, i.e., the nature of the fixed points is independent of the opinion update rules. This is particularly useful given few empirical results for the precise details of the rules employed by real agents~\cite{Fernandez-Gracia:2014}. 
The speed of the dynamics near the polarized state does depend on the form of update rules and we consider the consensus time $T_C = \int_{0}^{1} (\dot m)^{-1}dm$, the time taken by the system to reach consensus from the polarized state.  In quenched Ising systems this quantity is known as the saturation time. To circumvent the divergences at the fixed points (where $\dot m =0$), we note that the global opinion evolves in a step of $2/N$ in a finite system and alter the integration limits accordingly, $T_C \sim \lim_{N\to\infty} \int_{2a/N}^{1-2b/N} \frac{dm}{\dot m}$. 
Here $a$ and $b$ denote small positive integers which represent the first and final few steps where the mean field picture breaks down and the microscopic (stochastic) details are important. 
Factoring out the logarithmic divergence, we have 
\begin{align}
\frac{T_C}{\ln N}\, \sim \,\left( \frac{dm}{d\dot m}\right)_{m=0} +\left( - \frac{dm}{d\dot m}\right)_{m=1}.
\label{eq:TC}
\end{align}
This result holds if and only if there is no fixed point at an intermediate global opinion and both terms on the \textit{r.h.s.} are positive. 
Physically the first term measures depolarizing time and the second consensus approaching time. 
From Eq.~\eqref{eq:dmdotdm1}, we see that the consensus approaching time decreases and saturates as bias becomes stronger. 
In contrast we see from Eq.~\eqref{eq:dmdotdm0} that the depolarising time goes as $\eta$ for $\eta\gg1$. 
Therefor $T_C\sim\eta$ for $\eta\gg1$. 
%
%
%
%
%
%%%%%%%%%
\begin{figure}
\centering
\includegraphics[width=\linewidth]{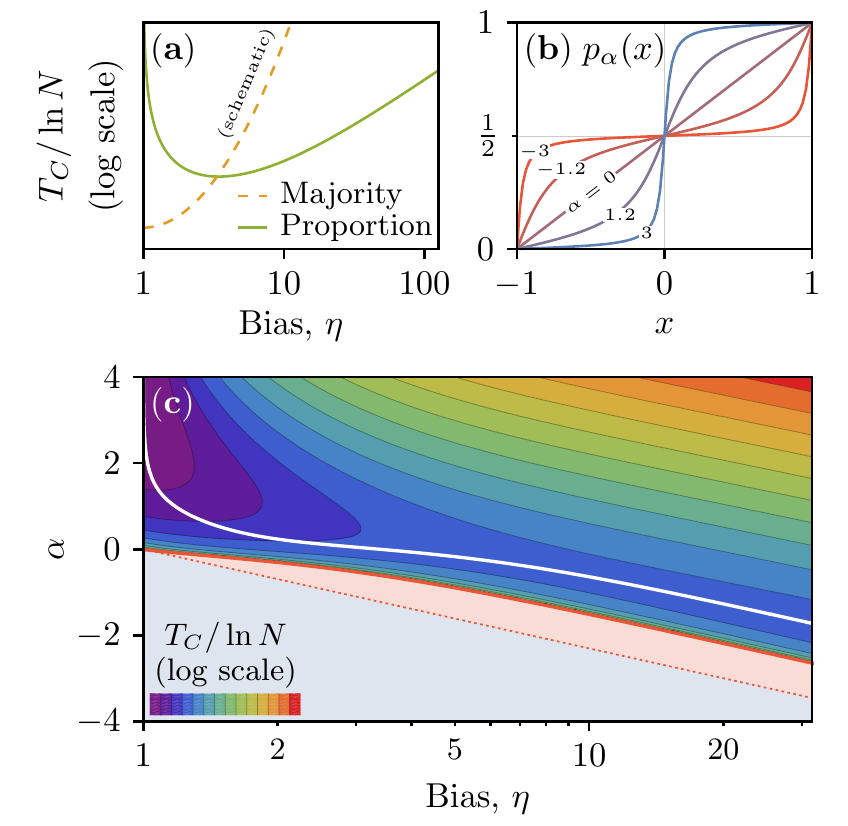}
\caption{
The time to consensus depends strongly on both opinion update rules and confirmation bias.
\textbf{(a)} 
The consensus time $T_C$ for majority-rule systems (dashed) increases with bias $\eta$ whereas for for proportion-rule systems (solid), $T_C$ clearly shows non-monotonic dependence. 
\textbf{(b)} 
Family of opinion update rules $p_\alpha(x)$ (exact form in the text) so that rules with $\alpha>0$  type-I and those with $\alpha<0$ are type-II.  The proportion and majority rules correspond to $\alpha = 0$ and $\alpha =\infty$, respectively.
\textbf{(c)} 
For all rules, there exists a minimum in $T_C$ at some $\eta>1$ marked by the white curve. 
For type-II rules ($\alpha<0$), we observe three regions (in the order of increasing bias): where the polarised state is the only attractive fixed point (grey shading); where consensus is also attractive (red shading); and where consensus prevail as the only attractive fixed point (colour scale). 
The minimum in $T_C$ exists in the final region. 
\label{fig:TC}
}
\end{figure}
%%%%%%%%

In Fig.~\ref{fig:TC}(a), we sketch the time to consensus in our model for the two well-known update rules: the majority rule and the proportion rule (voter model). Remarkably, for the proportion rule, $T_C$ develops a minimum at an intermediate bias, $\eta>1$ and this suggests that there could be an optimal bias for fast consensus formation.  Such non-monotonic behavior is absent with the majority rule where $T_C$ simply increases with $\eta$.  To systematically explore a family of update rules, we adopt a functional parameterization $p_\alpha(x) = \frac{1}{2}+\frac{1}{\pi}\tan^{-1}(e^\alpha \tan\frac{x\pi}{2})$ as shown in Fig.~\ref{fig:TC}(b) so that $p_\alpha(x)$ is type-I if $\alpha>0$ and type-II if $\alpha<0$.  The proportion rule corresponds to $\alpha=0$ and the majority rule to $\alpha = \infty$.

Figure~\ref{fig:TC}(c) illustrates how $T_C$ depends on both the update rules $\alpha$ and the bias $\eta$. 
For type-II rules ($\alpha<0$) and $\eta=1$ (no bias), the polarized state $m=0$ is the only attractive fixed point (grey shading).
As $\eta$ increases, first the consensus state is stabilized (red shading) and then the polarized state is destabilized (color scale) and the system flows uninterrupted from $m=0$ to $m=\pm1$ with $T_C$ given by Eq.~\eqref{eq:TC}.  
The minimum of $T_C$ shown previously for the proportion rule is now seen to be quite general for all rules. 
The initial decrease in $T_C$ results from disproportionate decreases in forward (minority-to-majority) and backward (majority-to-minority) conversions. 
Though all agents find themselves in neighborhoods of agreement, majority agents form more agreeing neighborhoods since it is easier to find like-minded agents. 
In consequence although the rate of both forward and backward conversions decrease with increasing bias, backward conversions decrease more than forward conversions; hence, the speed up of consensus formation. The increase in $T_C$ at larger $\eta$ results from the fact that the likelihood of a backward conversion cannot drop below zero but backward conversions continue to decrease; therefore, fewer net conversions and longer $T_C$. 

To further elucidate the impact of bias in our opinion model, we consider an extreme case where a fair coin toss decides the outcome of each opinion update, i.e., $p(x)=1/2$. 
In this case, the majority opinion suffers net outward conversions since inbound conversions originate from a smaller population and thus are less frequent than outward conversions. Hence, equal mixing of opinions is stable. 
Social interactions, as encapsulated in opinion update rules, can however change this picture. The proportion rule, for example, suppresses the probability of a majority-to-minority conversion and enhances the reverse in such a balanced way that each opinion population does not grow or shrink regardless of the global opinion. The majority rule reinforces this effect even further and, as a result, the only conversions left in the system are from minority to majority. 

When subject to bias, individuals now form a distorted worldview as they embed themselves among kindred spirits. 
Agents in the majority overestimate their majority and consequently rarely switch sides. 
Similarly, those in the minority are less likely to switch but their social circles are also less likely to be homogeneous, meaning conversions to the majority, though less frequent with bias, are still present. In short, intermediate confirmation bias can result in an increase in net conversions from minority to majority, hence speeding up agreement formation, but strong bias deprives the system of any conversions, resulting in longer consensus time.

 A number of generalizations of our model are immediately apparent. When consensus is stable, as for strong bias, a small perturbation in global opinion cannot lead to a permanent global opinion shift, a result at odds with how new, sparely known ideas can supersede old beliefs and become common knowledge. In our model, a natural way to capture this phenomenon is to break opinion symmetry, making one opinion more stable than the other, as with Ising spins in an external field. 
Such an asymmetry may also capture other social spreading phenomena such as the propagation of rumors or diseases (see, e.g., Ref.~\cite{Castellano:2003}).

The choice of update rules poses an important challenge to opinion dynamics modeling, particularly in the absence of strong empirical constraints.  One approach is to rationalize agents so that their update dynamics are a result of individual optimizations, such as fastest consensus formation or longest time with the opinion on which the society reaches consensus. With clear individual objectives, each received signal is carefully interpreted and optimal strategy deduced. 
It would be very interesting to see whether heuristic opinion update rules, such as in our work, could emerge in a system of rational individuals performing sophisticated calculations, e.g., using Bayesian inference (see, Ref.~\cite{Acemoglu:2011} for a review).

Insights into social dynamics could also allow large-scale opinion engineering.  Indeed, the idea that collective beliefs can be manipulated is not new; for example, see Ref.~\cite{DellaVigna:2007} where it was empirically shown that the arrival of The Fox News affected the Republican vote share. As an illustration, our results suggest that the speed of collective opinion formation could be controlled if one can tune individual dynamics.  In online social networks such as Facebook, algorithms exist such that interactions determine information flows between two users; hence, amplifying confirmation bias. Certainly a reverse algorithm could be engineered and this could have important implications for mass opinion manipulations. For example, an authoritarian regime might want to slow down collective opinion formation to allow more time to steer the dynamics in the preferred direction before speeding up consensus formation. 

Social regularities, such as common culture or languages, and their spontaneous emergence, show similarities to collective behavior in physical systems, thus raising the hope that some social phenomena could be understood in the context of models and concepts from statistical physics. Just as magnetism rests on the microscopic dynamics of spin, collective social phenomena arises from human behaviors.  While we have focused here on confirmation bias, increasingly quantitative approaches in behavioral sciences are offer important new ingredients to understanding collective social phenomena.  By incorporating such laws in simple model systems, we expect further advances in the physics of social networks.

This work was supported in part by funding from the Okinawa Institute of Science and Technology Graduate University.

\bibliographystyle{apsrev4-1_PRX_style} % tweaked bibtex style file to show the titles of cited articles
\bibliography{biblo}

\begin{thebibliography}{42}%
\makeatletter
\providecommand \@ifxundefined [1]{%
 \@ifx{#1\undefined}
}%
\providecommand \@ifnum [1]{%
 \ifnum #1\expandafter \@firstoftwo
 \else \expandafter \@secondoftwo
 \fi
}%
\providecommand \@ifx [1]{%
 \ifx #1\expandafter \@firstoftwo
 \else \expandafter \@secondoftwo
 \fi
}%
\providecommand \natexlab [1]{#1}%
\providecommand \enquote  [1]{``#1''}%
\providecommand \bibnamefont  [1]{#1}%
\providecommand \bibfnamefont [1]{#1}%
\providecommand \citenamefont [1]{#1}%
\providecommand \href@noop [0]{\@secondoftwo}%
\providecommand \href [0]{\begingroup \@sanitize@url \@href}%
\providecommand \@href[1]{\@@startlink{#1}\@@href}%
\providecommand \@@href[1]{\endgroup#1\@@endlink}%
\providecommand \@sanitize@url [0]{\catcode `\\12\catcode `\$12\catcode
  `\&12\catcode `\#12\catcode `\^12\catcode `\_12\catcode `\%12\relax}%
\providecommand \@@startlink[1]{}%
\providecommand \@@endlink[0]{}%
\providecommand \url  [0]{\begingroup\@sanitize@url \@url }%
\providecommand \@url [1]{\endgroup\@href {#1}{\urlprefix }}%
\providecommand \urlprefix  [0]{URL }%
\providecommand \Eprint [0]{\href }%
\providecommand \doibase [0]{http://dx.doi.org/}%
\providecommand \selectlanguage [0]{\@gobble}%
\providecommand \bibinfo  [0]{\@secondoftwo}%
\providecommand \bibfield  [0]{\@secondoftwo}%
\providecommand \translation [1]{[#1]}%
\providecommand \BibitemOpen [0]{}%
\providecommand \bibitemStop [0]{}%
\providecommand \bibitemNoStop [0]{.\EOS\space}%
\providecommand \EOS [0]{\spacefactor3000\relax}%
\providecommand \BibitemShut  [1]{\csname bibitem#1\endcsname}%
\let\auto@bib@innerbib\@empty
%</preamble>
\bibitem [{gay(ly 4)}]{gaymarriage}%
  \BibitemOpen
  \href
  {http://www.economist.com/news/united-states/21656667-nine-judges-are-being-asked-compensate-political-stalemate-both-troubling}
  {\bibfield  {title} {\emph {\bibinfo {title} {Change is gonna come},\
  }}}\bibinfo {howpublished} {{{The Economist}}} (\bibinfo {year} {2015, July
  4})\BibitemShut {NoStop}%
\bibitem [{\citenamefont {Acemoglu}\ and\ \citenamefont
  {Ozdaglar}(2011)}]{Acemoglu:2011}%
  \BibitemOpen
  \bibfield  {author} {\bibinfo {author} {\bibfnamefont {D.}~\bibnamefont
  {Acemoglu}}\ and\ \bibinfo {author} {\bibfnamefont {A.}~\bibnamefont
  {Ozdaglar}},\ }\bibfield  {Opinion Dynamics and Learning in Social Networks}
  {\emph {\bibinfo {title} {Opinion dynamics and learning in social networks},\
  }}\href {http://dx.doi.org/10.1007/s13235-010-0004-1} {\bibfield  {journal}
  {\bibinfo  {journal} {Dynamic Games and Applications}\ }\textbf {\bibinfo
  {volume} {1}},\ \bibinfo {pages} {3} (\bibinfo {year} {2011})}\BibitemShut
  {NoStop}%
\bibitem [{\citenamefont {Tversky}\ and\ \citenamefont
  {Kahneman}(1974)}]{Tversky:1974}%
  \BibitemOpen
  \bibfield  {author} {\bibinfo {author} {\bibfnamefont {A.}~\bibnamefont
  {Tversky}}\ and\ \bibinfo {author} {\bibfnamefont {D.}~\bibnamefont
  {Kahneman}},\ }\bibfield  {Judgment under Uncertainty: Heuristics and Biases}
  {\emph {\bibinfo {title} {Judgment under uncertainty: Heuristics and
  biases},\ }}\href {http://www.sciencemag.org/content/185/4157/1124.abstract}
  {\bibfield  {journal} {\bibinfo  {journal} {Science}\ }\textbf {\bibinfo
  {volume} {185}},\ \bibinfo {pages} {1124} (\bibinfo {year}
  {1974})}\BibitemShut {NoStop}%
\bibitem [{\citenamefont {Nickerson}(1998)}]{Nickerson:1998}%
  \BibitemOpen
  \bibfield  {author} {\bibinfo {author} {\bibfnamefont {R.~S.}\ \bibnamefont
  {Nickerson}},\ }\bibfield  {Confirmation Bias: A Ubiquitous Phenomenon in
  Many Guises} {\emph {\bibinfo {title} {Confirmation bias: A ubiquitous
  phenomenon in many guises},\ }}\href@noop {} {\bibfield  {journal} {\bibinfo
  {journal} {Review of General Psychology}\ }\textbf {\bibinfo {volume} {2}},\
  \bibinfo {pages} {175} (\bibinfo {year} {1998})}\BibitemShut {NoStop}%
\bibitem [{\citenamefont {Rosenblat}\ and\ \citenamefont
  {Mobius}(2004)}]{Rosenblat:2004}%
  \BibitemOpen
  \bibfield  {author} {\bibinfo {author} {\bibfnamefont {T.~S.}\ \bibnamefont
  {Rosenblat}}\ and\ \bibinfo {author} {\bibfnamefont {M.~M.}\ \bibnamefont
  {Mobius}},\ }\bibfield  {Getting Closer or Drifting Apart?} {\emph {\bibinfo
  {title} {Getting closer or drifting apart?}\ }}\href
  {http://qje.oxfordjournals.org/content/119/3/971.abstract} {\bibfield
  {journal} {\bibinfo  {journal} {The Quarterly Journal of Economics}\ }\textbf
  {\bibinfo {volume} {119}},\ \bibinfo {pages} {971} (\bibinfo {year}
  {2004})}\BibitemShut {NoStop}%
\bibitem [{\citenamefont {Stroud}(2008)}]{Stroud:2008}%
  \BibitemOpen
  \bibfield  {author} {\bibinfo {author} {\bibfnamefont {N.~J.}\ \bibnamefont
  {Stroud}},\ }\bibfield  {Media Use and Political Predispositions: Revisiting
  the Concept of Selective Exposure} {\emph {\bibinfo {title} {Media use and
  political predispositions: Revisiting the concept of selective exposure},\
  }}\href@noop {} {\bibfield  {journal} {\bibinfo  {journal} {Political
  Behavior}\ }\textbf {\bibinfo {volume} {30}},\ \bibinfo {pages} {341}
  (\bibinfo {year} {2008})}\BibitemShut {NoStop}%
\bibitem [{\citenamefont {Valentino}\ \emph {et~al.}(2009)\citenamefont
  {Valentino}, \citenamefont {Banks}, \citenamefont {Hutchings},\ and\
  \citenamefont {Davis}}]{Valentino:2009}%
  \BibitemOpen
  \bibfield  {author} {\bibinfo {author} {\bibfnamefont {N.~A.}\ \bibnamefont
  {Valentino}}, \bibinfo {author} {\bibfnamefont {A.~J.}\ \bibnamefont
  {Banks}}, \bibinfo {author} {\bibfnamefont {V.~L.}\ \bibnamefont
  {Hutchings}}, \ and\ \bibinfo {author} {\bibfnamefont {A.~K.}\ \bibnamefont
  {Davis}},\ }\bibfield  {Selective Exposure in the Internet Age: The
  Interaction between Anxiety and Information Utility} {\emph {\bibinfo {title}
  {Selective exposure in the internet age: The interaction between anxiety and
  information utility},\ }}\href
  {http://dx.doi.org/10.1111/j.1467-9221.2009.00716.x} {\bibfield  {journal}
  {\bibinfo  {journal} {Political Psychology}\ }\textbf {\bibinfo {volume}
  {30}},\ \bibinfo {pages} {591} (\bibinfo {year} {2009})}\BibitemShut
  {NoStop}%
\bibitem [{\citenamefont {Garrett}(2009)}]{Garrett:2009}%
  \BibitemOpen
  \bibfield  {author} {\bibinfo {author} {\bibfnamefont {R.~K.}\ \bibnamefont
  {Garrett}},\ }\bibfield  {Politically Motivated Reinforcement Seeking:
  Reframing the Selective Exposure Debate} {\emph {\bibinfo {title}
  {Politically motivated reinforcement seeking: Reframing the selective
  exposure debate},\ }}\href
  {http://dx.doi.org/10.1111/j.1460-2466.2009.01452.x} {\bibfield  {journal}
  {\bibinfo  {journal} {Journal of Communication}\ }\textbf {\bibinfo {volume}
  {59}},\ \bibinfo {pages} {676} (\bibinfo {year} {2009})}\BibitemShut
  {NoStop}%
\bibitem [{\citenamefont {Bakshy}\ \emph {et~al.}(2015)\citenamefont {Bakshy},
  \citenamefont {Messing},\ and\ \citenamefont {Adamic}}]{Bakshy:2015}%
  \BibitemOpen
  \bibfield  {author} {\bibinfo {author} {\bibfnamefont {E.}~\bibnamefont
  {Bakshy}}, \bibinfo {author} {\bibfnamefont {S.}~\bibnamefont {Messing}}, \
  and\ \bibinfo {author} {\bibfnamefont {L.~A.}\ \bibnamefont {Adamic}},\
  }\bibfield  {Exposure to ideologically diverse news and opinion on Facebook}
  {\emph {\bibinfo {title} {Exposure to ideologically diverse news and opinion
  on facebook},\ }}\href
  {http://www.sciencemag.org/content/348/6239/1130.abstract} {\bibfield
  {journal} {\bibinfo  {journal} {Science}\ }\textbf {\bibinfo {volume}
  {348}},\ \bibinfo {pages} {1130} (\bibinfo {year} {2015})}\BibitemShut
  {NoStop}%
\bibitem [{\citenamefont {Frey}(1986)}]{Frey:1986}%
  \BibitemOpen
  \bibfield  {author} {\bibinfo {author} {\bibfnamefont {D.}~\bibnamefont
  {Frey}},\ }\bibfield  {Recent research on selective exposure to information}
  {\emph {\bibinfo {title} {Recent research on selective exposure to
  information},\ }}\href@noop {} {\bibfield  {journal} {\bibinfo  {journal}
  {Advances in experimental social psychology}\ }\textbf {\bibinfo {volume}
  {19}},\ \bibinfo {pages} {41} (\bibinfo {year} {1986})}\BibitemShut {NoStop}%
\bibitem [{\citenamefont {Lord}\ \emph {et~al.}(1979)\citenamefont {Lord},
  \citenamefont {Ross},\ and\ \citenamefont {Lepper}}]{Lord:1979}%
  \BibitemOpen
  \bibfield  {author} {\bibinfo {author} {\bibfnamefont {C.~G.}\ \bibnamefont
  {Lord}}, \bibinfo {author} {\bibfnamefont {L.}~\bibnamefont {Ross}}, \ and\
  \bibinfo {author} {\bibfnamefont {M.~R.}\ \bibnamefont {Lepper}},\ }\bibfield
   {Biased assimilation and attitude polarization: the effects of prior
  theories on subsequently considered evidence.} {\emph {\bibinfo {title}
  {Biased assimilation and attitude polarization: the effects of prior theories
  on subsequently considered evidence.}\ }}\href@noop {} {\bibfield  {journal}
  {\bibinfo  {journal} {Journal of personality and social psychology}\ }\textbf
  {\bibinfo {volume} {37}},\ \bibinfo {pages} {2098} (\bibinfo {year}
  {1979})}\BibitemShut {NoStop}%
\bibitem [{\citenamefont {Griffin}\ and\ \citenamefont
  {Tversky}(1992)}]{Griffin:1992}%
  \BibitemOpen
  \bibfield  {author} {\bibinfo {author} {\bibfnamefont {D.}~\bibnamefont
  {Griffin}}\ and\ \bibinfo {author} {\bibfnamefont {A.}~\bibnamefont
  {Tversky}},\ }\bibfield  {The weighing of evidence and the determinants of
  confidence} {\emph {\bibinfo {title} {The weighing of evidence and the
  determinants of confidence},\ }}\href
  {http://www.sciencedirect.com/science/article/pii/001002859290013R}
  {\bibfield  {journal} {\bibinfo  {journal} {Cognitive Psychology}\ }\textbf
  {\bibinfo {volume} {24}},\ \bibinfo {pages} {411 } (\bibinfo {year}
  {1992})}\BibitemShut {NoStop}%
\bibitem [{\citenamefont {Galam}\ \emph {et~al.}(1982)\citenamefont {Galam},
  \citenamefont {Feigenblat},\ and\ \citenamefont {Shapir}}]{Galam:1982}%
  \BibitemOpen
  \bibfield  {author} {\bibinfo {author} {\bibfnamefont {S.}~\bibnamefont
  {Galam}}, \bibinfo {author} {\bibfnamefont {Y.~G.}\ \bibnamefont
  {Feigenblat}}, \ and\ \bibinfo {author} {\bibfnamefont {Y.}~\bibnamefont
  {Shapir}},\ }\bibfield  {Sociophysics: A new approach of sociological
  collective behaviour. I. mean‐behaviour description of a strike} {\emph
  {\bibinfo {title} {Sociophysics: A new approach of sociological collective
  behaviour. i. mean‐behaviour description of a strike},\ }}\href
  {http://dx.doi.org/10.1080/0022250X.1982.9989929} {\bibfield  {journal}
  {\bibinfo  {journal} {The Journal of Mathematical Sociology}\ }\textbf
  {\bibinfo {volume} {9}},\ \bibinfo {pages} {1} (\bibinfo {year}
  {1982})}\BibitemShut {NoStop}%
\bibitem [{\citenamefont {Galam}\ and\ \citenamefont
  {Moscovici}(1991)}]{Galam:1991}%
  \BibitemOpen
  \bibfield  {author} {\bibinfo {author} {\bibfnamefont {S.}~\bibnamefont
  {Galam}}\ and\ \bibinfo {author} {\bibfnamefont {S.}~\bibnamefont
  {Moscovici}},\ }\bibfield  {Towards a theory of collective phenomena:
  Consensus and attitude changes in groups} {\emph {\bibinfo {title} {Towards a
  theory of collective phenomena: Consensus and attitude changes in groups},\
  }}\href {http://dx.doi.org/10.1002/ejsp.2420210105} {\bibfield  {journal}
  {\bibinfo  {journal} {European Journal of Social Psychology}\ }\textbf
  {\bibinfo {volume} {21}},\ \bibinfo {pages} {49} (\bibinfo {year}
  {1991})}\BibitemShut {NoStop}%
\bibitem [{\citenamefont {Deffuant}\ \emph {et~al.}(2000)\citenamefont
  {Deffuant}, \citenamefont {Neau}, \citenamefont {Amblard},\ and\
  \citenamefont {Weisbuch}}]{Deffuant:2000}%
  \BibitemOpen
  \bibfield  {author} {\bibinfo {author} {\bibfnamefont {G.}~\bibnamefont
  {Deffuant}}, \bibinfo {author} {\bibfnamefont {D.}~\bibnamefont {Neau}},
  \bibinfo {author} {\bibfnamefont {F.}~\bibnamefont {Amblard}}, \ and\
  \bibinfo {author} {\bibfnamefont {G.}~\bibnamefont {Weisbuch}},\ }\bibfield
  {Mixing beliefs among interacting agents} {\emph {\bibinfo {title} {Mixing
  beliefs among interacting agents},\ }}\href
  {http://www.worldscientific.com/doi/abs/10.1142/S0219525900000078} {\bibfield
   {journal} {\bibinfo  {journal} {Advances in Complex Systems}\ }\textbf
  {\bibinfo {volume} {03}},\ \bibinfo {pages} {87} (\bibinfo {year}
  {2000})}\BibitemShut {NoStop}%
\bibitem [{\citenamefont {Krapivsky}\ and\ \citenamefont
  {Redner}(2003)}]{Krapivsky:2003}%
  \BibitemOpen
  \bibfield  {author} {\bibinfo {author} {\bibfnamefont {P.~L.}\ \bibnamefont
  {Krapivsky}}\ and\ \bibinfo {author} {\bibfnamefont {S.}~\bibnamefont
  {Redner}},\ }\bibfield  {Dynamics of Majority Rule in Two-State Interacting
  Spin Systems} {\emph {\bibinfo {title} {Dynamics of majority rule in
  two-state interacting spin systems},\ }}\href
  {http://link.aps.org/doi/10.1103/PhysRevLett.90.238701} {\bibfield  {journal}
  {\bibinfo  {journal} {Phys. Rev. Lett.}\ }\textbf {\bibinfo {volume} {90}},\
  \bibinfo {pages} {238701} (\bibinfo {year} {2003})}\BibitemShut {NoStop}%
\bibitem [{\citenamefont {Mobilia}(2003)}]{Mobilia:2003}%
  \BibitemOpen
  \bibfield  {author} {\bibinfo {author} {\bibfnamefont {M.}~\bibnamefont
  {Mobilia}},\ }\bibfield  {Does a Single Zealot Affect an Infinite Group of
  Voters?} {\emph {\bibinfo {title} {Does a single zealot affect an infinite
  group of voters?}\ }}\href
  {http://link.aps.org/doi/10.1103/PhysRevLett.91.028701} {\bibfield  {journal}
  {\bibinfo  {journal} {Phys. Rev. Lett.}\ }\textbf {\bibinfo {volume} {91}},\
  \bibinfo {pages} {028701} (\bibinfo {year} {2003})}\BibitemShut {NoStop}%
\bibitem [{\citenamefont {Sood}\ and\ \citenamefont
  {Redner}(2005)}]{Sood:2005}%
  \BibitemOpen
  \bibfield  {author} {\bibinfo {author} {\bibfnamefont {V.}~\bibnamefont
  {Sood}}\ and\ \bibinfo {author} {\bibfnamefont {S.}~\bibnamefont {Redner}},\
  }\bibfield  {Voter Model on Heterogeneous Graphs} {\emph {\bibinfo {title}
  {Voter model on heterogeneous graphs},\ }}\href
  {http://link.aps.org/doi/10.1103/PhysRevLett.94.178701} {\bibfield  {journal}
  {\bibinfo  {journal} {Phys. Rev. Lett.}\ }\textbf {\bibinfo {volume} {94}},\
  \bibinfo {pages} {178701} (\bibinfo {year} {2005})}\BibitemShut {NoStop}%
\bibitem [{\citenamefont {Suchecki}\ \emph {et~al.}(2005)\citenamefont
  {Suchecki}, \citenamefont {Egu{\'\i}luz},\ and\ \citenamefont
  {Miguel}}]{Suchecki:2005}%
  \BibitemOpen
  \bibfield  {author} {\bibinfo {author} {\bibfnamefont {K.}~\bibnamefont
  {Suchecki}}, \bibinfo {author} {\bibfnamefont {V.~M.}\ \bibnamefont
  {Egu{\'\i}luz}}, \ and\ \bibinfo {author} {\bibfnamefont {M.~S.}\
  \bibnamefont {Miguel}},\ }\bibfield  {Conservation laws for the voter model
  in complex networks} {\emph {\bibinfo {title} {Conservation laws for the
  voter model in complex networks},\ }}\href
  {http://stacks.iop.org/0295-5075/69/i=2/a=228} {\bibfield  {journal}
  {\bibinfo  {journal} {EPL (Europhysics Letters)}\ }\textbf {\bibinfo {volume}
  {69}},\ \bibinfo {pages} {228} (\bibinfo {year} {2005})}\BibitemShut
  {NoStop}%
\bibitem [{\citenamefont {Castellano}\ \emph {et~al.}(2009)\citenamefont
  {Castellano}, \citenamefont {Fortunato},\ and\ \citenamefont
  {Loreto}}]{Castellano:2009}%
  \BibitemOpen
  \bibfield  {author} {\bibinfo {author} {\bibfnamefont {C.}~\bibnamefont
  {Castellano}}, \bibinfo {author} {\bibfnamefont {S.}~\bibnamefont
  {Fortunato}}, \ and\ \bibinfo {author} {\bibfnamefont {V.}~\bibnamefont
  {Loreto}},\ }\bibfield  {Statistical physics of social dynamics} {\emph
  {\bibinfo {title} {Statistical physics of social dynamics},\ }}\href
  {http://link.aps.org/doi/10.1103/RevModPhys.81.591} {\bibfield  {journal}
  {\bibinfo  {journal} {Rev. Mod. Phys.}\ }\textbf {\bibinfo {volume} {81}},\
  \bibinfo {pages} {591} (\bibinfo {year} {2009})}\BibitemShut {NoStop}%
\bibitem [{\citenamefont {Clifford}\ and\ \citenamefont
  {Sudbury}(1973)}]{Clifford:1973}%
  \BibitemOpen
  \bibfield  {author} {\bibinfo {author} {\bibfnamefont {P.}~\bibnamefont
  {Clifford}}\ and\ \bibinfo {author} {\bibfnamefont {A.}~\bibnamefont
  {Sudbury}},\ }\bibfield  {A model for spatial conflict} {\emph {\bibinfo
  {title} {A model for spatial conflict},\ }}\href
  {http://biomet.oxfordjournals.org/content/60/3/581.abstract} {\bibfield
  {journal} {\bibinfo  {journal} {Biometrika}\ }\textbf {\bibinfo {volume}
  {60}},\ \bibinfo {pages} {581} (\bibinfo {year} {1973})}\BibitemShut
  {NoStop}%
\bibitem [{\citenamefont {Holley}\ and\ \citenamefont
  {Liggett}(1975)}]{Holley:1975}%
  \BibitemOpen
  \bibfield  {author} {\bibinfo {author} {\bibfnamefont {R.~A.}\ \bibnamefont
  {Holley}}\ and\ \bibinfo {author} {\bibfnamefont {T.~M.}\ \bibnamefont
  {Liggett}},\ }\bibfield  {Ergodic Theorems for Weakly Interacting Infinite
  Systems and the Voter Model} {\emph {\bibinfo {title} {Ergodic theorems for
  weakly interacting infinite systems and the voter model},\ }}\href
  {http://www.jstor.org/stable/2959329} {\bibfield  {journal} {\bibinfo
  {journal} {The Annals of Probability}\ }\textbf {\bibinfo {volume} {3}},\
  \bibinfo {pages} {643} (\bibinfo {year} {1975})}\BibitemShut {NoStop}%
\bibitem [{\citenamefont {Krapivsky}\ \emph {et~al.}(2010)\citenamefont
  {Krapivsky}, \citenamefont {Redner},\ and\ \citenamefont
  {Ben-Naim}}]{Krapivsky:2010}%
  \BibitemOpen
  \bibfield  {author} {\bibinfo {author} {\bibfnamefont {P.~L.}\ \bibnamefont
  {Krapivsky}}, \bibinfo {author} {\bibfnamefont {S.}~\bibnamefont {Redner}}, \
  and\ \bibinfo {author} {\bibfnamefont {E.}~\bibnamefont {Ben-Naim}},\
  }\href@noop {} {\emph {\bibinfo {title} {A kinetic view of statistical
  physics}}}\ (\bibinfo  {publisher} {Cambridge University Press},\ \bibinfo
  {year} {2010})\BibitemShut {NoStop}%
\bibitem [{\citenamefont {Castellano}\ \emph {et~al.}(2003)\citenamefont
  {Castellano}, \citenamefont {Vilone},\ and\ \citenamefont
  {Vespignani}}]{Castellano:2003}%
  \BibitemOpen
  \bibfield  {author} {\bibinfo {author} {\bibfnamefont {C.}~\bibnamefont
  {Castellano}}, \bibinfo {author} {\bibfnamefont {D.}~\bibnamefont {Vilone}},
  \ and\ \bibinfo {author} {\bibfnamefont {A.}~\bibnamefont {Vespignani}},\
  }\bibfield  {Incomplete ordering of the voter model on small-world networks}
  {\emph {\bibinfo {title} {Incomplete ordering of the voter model on
  small-world networks},\ }}\href
  {http://stacks.iop.org/0295-5075/63/i=1/a=153} {\bibfield  {journal}
  {\bibinfo  {journal} {EPL (Europhysics Letters)}\ }\textbf {\bibinfo {volume}
  {63}},\ \bibinfo {pages} {153} (\bibinfo {year} {2003})}\BibitemShut
  {NoStop}%
\bibitem [{\citenamefont {Vilone}\ and\ \citenamefont
  {Castellano}(2004)}]{Vilone:2004}%
  \BibitemOpen
  \bibfield  {author} {\bibinfo {author} {\bibfnamefont {D.}~\bibnamefont
  {Vilone}}\ and\ \bibinfo {author} {\bibfnamefont {C.}~\bibnamefont
  {Castellano}},\ }\bibfield  {Solution of voter model dynamics on annealed
  small-world networks} {\emph {\bibinfo {title} {Solution of voter model
  dynamics on annealed small-world networks},\ }}\href
  {http://link.aps.org/doi/10.1103/PhysRevE.69.016109} {\bibfield  {journal}
  {\bibinfo  {journal} {Phys. Rev. E}\ }\textbf {\bibinfo {volume} {69}},\
  \bibinfo {pages} {016109} (\bibinfo {year} {2004})}\BibitemShut {NoStop}%
\bibitem [{\citenamefont {Sood}\ \emph {et~al.}(2008)\citenamefont {Sood},
  \citenamefont {Antal},\ and\ \citenamefont {Redner}}]{Sood:2008}%
  \BibitemOpen
  \bibfield  {author} {\bibinfo {author} {\bibfnamefont {V.}~\bibnamefont
  {Sood}}, \bibinfo {author} {\bibfnamefont {T.}~\bibnamefont {Antal}}, \ and\
  \bibinfo {author} {\bibfnamefont {S.}~\bibnamefont {Redner}},\ }\bibfield
  {Voter models on heterogeneous networks} {\emph {\bibinfo {title} {Voter
  models on heterogeneous networks},\ }}\href
  {http://link.aps.org/doi/10.1103/PhysRevE.77.041121} {\bibfield  {journal}
  {\bibinfo  {journal} {Phys. Rev. E}\ }\textbf {\bibinfo {volume} {77}},\
  \bibinfo {pages} {041121} (\bibinfo {year} {2008})}\BibitemShut {NoStop}%
\bibitem [{\citenamefont {Holme}\ and\ \citenamefont
  {Newman}(2006)}]{Holme:2006}%
  \BibitemOpen
  \bibfield  {author} {\bibinfo {author} {\bibfnamefont {P.}~\bibnamefont
  {Holme}}\ and\ \bibinfo {author} {\bibfnamefont {M.~E.~J.}\ \bibnamefont
  {Newman}},\ }\bibfield  {Nonequilibrium phase transition in the coevolution
  of networks and opinions} {\emph {\bibinfo {title} {Nonequilibrium phase
  transition in the coevolution of networks and opinions},\ }}\href
  {http://link.aps.org/doi/10.1103/PhysRevE.74.056108} {\bibfield  {journal}
  {\bibinfo  {journal} {Phys. Rev. E}\ }\textbf {\bibinfo {volume} {74}},\
  \bibinfo {pages} {056108} (\bibinfo {year} {2006})}\BibitemShut {NoStop}%
\bibitem [{\citenamefont {Benczik}\ \emph {et~al.}(2009)\citenamefont
  {Benczik}, \citenamefont {Benczik}, \citenamefont {Schmittmann},\ and\
  \citenamefont {Zia}}]{Benczik:2009}%
  \BibitemOpen
  \bibfield  {author} {\bibinfo {author} {\bibfnamefont {I.~J.}\ \bibnamefont
  {Benczik}}, \bibinfo {author} {\bibfnamefont {S.~Z.}\ \bibnamefont
  {Benczik}}, \bibinfo {author} {\bibfnamefont {B.}~\bibnamefont
  {Schmittmann}}, \ and\ \bibinfo {author} {\bibfnamefont {R.~K.~P.}\
  \bibnamefont {Zia}},\ }\bibfield  {Opinion dynamics on an adaptive random
  network} {\emph {\bibinfo {title} {Opinion dynamics on an adaptive random
  network},\ }}\href {http://link.aps.org/doi/10.1103/PhysRevE.79.046104}
  {\bibfield  {journal} {\bibinfo  {journal} {Phys. Rev. E}\ }\textbf {\bibinfo
  {volume} {79}},\ \bibinfo {pages} {046104} (\bibinfo {year}
  {2009})}\BibitemShut {NoStop}%
\bibitem [{\citenamefont {Zschaler}\ \emph {et~al.}(2012)\citenamefont
  {Zschaler}, \citenamefont {B\"ohme}, \citenamefont {Sei\ss{}inger},
  \citenamefont {Huepe},\ and\ \citenamefont {Gross}}]{Zschaler:2012}%
  \BibitemOpen
  \bibfield  {author} {\bibinfo {author} {\bibfnamefont {G.}~\bibnamefont
  {Zschaler}}, \bibinfo {author} {\bibfnamefont {G.~A.}\ \bibnamefont
  {B\"ohme}}, \bibinfo {author} {\bibfnamefont {M.}~\bibnamefont
  {Sei\ss{}inger}}, \bibinfo {author} {\bibfnamefont {C.}~\bibnamefont
  {Huepe}}, \ and\ \bibinfo {author} {\bibfnamefont {T.}~\bibnamefont
  {Gross}},\ }\bibfield  {Early fragmentation in the adaptive voter model on
  directed networks} {\emph {\bibinfo {title} {Early fragmentation in the
  adaptive voter model on directed networks},\ }}\href
  {http://link.aps.org/doi/10.1103/PhysRevE.85.046107} {\bibfield  {journal}
  {\bibinfo  {journal} {Phys. Rev. E}\ }\textbf {\bibinfo {volume} {85}},\
  \bibinfo {pages} {046107} (\bibinfo {year} {2012})}\BibitemShut {NoStop}%
\bibitem [{\citenamefont {Durrett}\ \emph {et~al.}(2012)\citenamefont
  {Durrett}, \citenamefont {Gleeson}, \citenamefont {Lloyd}, \citenamefont
  {Mucha}, \citenamefont {Shi}, \citenamefont {Sivakoff}, \citenamefont
  {Socolar},\ and\ \citenamefont {Varghese}}]{Durrett:2012}%
  \BibitemOpen
  \bibfield  {author} {\bibinfo {author} {\bibfnamefont {R.}~\bibnamefont
  {Durrett}}, \bibinfo {author} {\bibfnamefont {J.~P.}\ \bibnamefont
  {Gleeson}}, \bibinfo {author} {\bibfnamefont {A.~L.}\ \bibnamefont {Lloyd}},
  \bibinfo {author} {\bibfnamefont {P.~J.}\ \bibnamefont {Mucha}}, \bibinfo
  {author} {\bibfnamefont {F.}~\bibnamefont {Shi}}, \bibinfo {author}
  {\bibfnamefont {D.}~\bibnamefont {Sivakoff}}, \bibinfo {author}
  {\bibfnamefont {J.~E.~S.}\ \bibnamefont {Socolar}}, \ and\ \bibinfo {author}
  {\bibfnamefont {C.}~\bibnamefont {Varghese}},\ }\bibfield  {Graph fission in
  an evolving voter model} {\emph {\bibinfo {title} {Graph fission in an
  evolving voter model},\ }}\href
  {http://www.pnas.org/content/109/10/3682.abstract} {\bibfield  {journal}
  {\bibinfo  {journal} {Proc. Nat. Acad. Sci.}\ }\textbf {\bibinfo {volume}
  {109}},\ \bibinfo {pages} {3682} (\bibinfo {year} {2012})}\BibitemShut
  {NoStop}%
\bibitem [{\citenamefont {Rogers}\ and\ \citenamefont
  {Gross}(2013)}]{Rogers:2013}%
  \BibitemOpen
  \bibfield  {author} {\bibinfo {author} {\bibfnamefont {T.}~\bibnamefont
  {Rogers}}\ and\ \bibinfo {author} {\bibfnamefont {T.}~\bibnamefont {Gross}},\
  }\bibfield  {Consensus time and conformity in the adaptive voter model}
  {\emph {\bibinfo {title} {Consensus time and conformity in the adaptive voter
  model},\ }}\href {http://link.aps.org/doi/10.1103/PhysRevE.88.030102}
  {\bibfield  {journal} {\bibinfo  {journal} {Phys. Rev. E}\ }\textbf {\bibinfo
  {volume} {88}},\ \bibinfo {pages} {030102} (\bibinfo {year}
  {2013})}\BibitemShut {NoStop}%
\bibitem [{\citenamefont {Adamic}\ and\ \citenamefont
  {Glance}(2005)}]{Adamic:2005}%
  \BibitemOpen
  \bibfield  {author} {\bibinfo {author} {\bibfnamefont {L.~A.}\ \bibnamefont
  {Adamic}}\ and\ \bibinfo {author} {\bibfnamefont {N.}~\bibnamefont
  {Glance}},\ }in\ \href {http://doi.acm.org/10.1145/1134271.1134277} {\emph
  {\bibinfo {booktitle} {Proceedings of the 3rd International Workshop on Link
  Discovery}}},\ \bibinfo {series and number} {LinkKDD '05}\ (\bibinfo
  {publisher} {ACM},\ \bibinfo {address} {New York, NY, USA},\ \bibinfo {year}
  {2005})\ pp.\ \bibinfo {pages} {36--43}\BibitemShut {NoStop}%
\bibitem [{\citenamefont {Lazarsfeld}\ and\ \citenamefont
  {Merton}(1954)}]{Lazarsfeld:1954}%
  \BibitemOpen
  \bibfield  {author} {\bibinfo {author} {\bibfnamefont {P.~F.}\ \bibnamefont
  {Lazarsfeld}}\ and\ \bibinfo {author} {\bibfnamefont {R.~K.}\ \bibnamefont
  {Merton}},\ }in\ \href@noop {} {\emph {\bibinfo {booktitle} {Freedom and
  control in modern society}}},\ \bibinfo {editor} {edited by\ \bibinfo
  {editor} {\bibfnamefont {M.}~\bibnamefont {Berger}}}\ (\bibinfo  {publisher}
  {New York: Van Nostrand},\ \bibinfo {year} {1954})\BibitemShut {NoStop}%
\bibitem [{\citenamefont {McPherson}\ \emph {et~al.}(2001)\citenamefont
  {McPherson}, \citenamefont {Smith-Lovin},\ and\ \citenamefont
  {Cook}}]{McPherson:2001}%
  \BibitemOpen
  \bibfield  {author} {\bibinfo {author} {\bibfnamefont {M.}~\bibnamefont
  {McPherson}}, \bibinfo {author} {\bibfnamefont {L.}~\bibnamefont
  {Smith-Lovin}}, \ and\ \bibinfo {author} {\bibfnamefont {J.~M.}\ \bibnamefont
  {Cook}},\ }\bibfield  {Birds of a Feather: Homophily in Social Networks}
  {\emph {\bibinfo {title} {Birds of a feather: Homophily in social networks},\
  }}\href {http://www.jstor.org/stable/2678628} {\bibfield  {journal} {\bibinfo
   {journal} {Annual Review of Sociology}\ }\textbf {\bibinfo {volume} {27}},\
  \bibinfo {pages} {415} (\bibinfo {year} {2001})}\BibitemShut {NoStop}%
\bibitem [{\citenamefont {DeMarzo}\ \emph {et~al.}(2003)\citenamefont
  {DeMarzo}, \citenamefont {Vayanos},\ and\ \citenamefont
  {Zwiebel}}]{DeMarzo:2003}%
  \BibitemOpen
  \bibfield  {author} {\bibinfo {author} {\bibfnamefont {P.~M.}\ \bibnamefont
  {DeMarzo}}, \bibinfo {author} {\bibfnamefont {D.}~\bibnamefont {Vayanos}}, \
  and\ \bibinfo {author} {\bibfnamefont {J.}~\bibnamefont {Zwiebel}},\
  }\bibfield  {Persuasion Bias, Social Influence, and Unidimensional Opinions}
  {\emph {\bibinfo {title} {Persuasion bias, social influence, and
  unidimensional opinions},\ }}\href
  {http://qje.oxfordjournals.org/content/118/3/909.abstract} {\bibfield
  {journal} {\bibinfo  {journal} {The Quarterly Journal of Economics}\ }\textbf
  {\bibinfo {volume} {118}},\ \bibinfo {pages} {909} (\bibinfo {year}
  {2003})}\BibitemShut {NoStop}%
\bibitem [{\citenamefont {de~Oliveira}\ \emph {et~al.}(1993)\citenamefont
  {de~Oliveira}, \citenamefont {Mendes},\ and\ \citenamefont
  {Santos}}]{Oliveira:1993}%
  \BibitemOpen
  \bibfield  {author} {\bibinfo {author} {\bibfnamefont {M.~J.}\ \bibnamefont
  {de~Oliveira}}, \bibinfo {author} {\bibfnamefont {J.~F.~F.}\ \bibnamefont
  {Mendes}}, \ and\ \bibinfo {author} {\bibfnamefont {M.~A.}\ \bibnamefont
  {Santos}},\ }\bibfield  {Nonequilibrium spin models with Ising universal
  behaviour} {\emph {\bibinfo {title} {Nonequilibrium spin models with ising
  universal behaviour},\ }}\href
  {http://stacks.iop.org/0305-4470/26/i=10/a=006} {\bibfield  {journal}
  {\bibinfo  {journal} {J. Phys. A}\ }\textbf {\bibinfo {volume} {26}},\
  \bibinfo {pages} {2317} (\bibinfo {year} {1993})}\BibitemShut {NoStop}%
\bibitem [{\citenamefont {Drouffe}\ and\ \citenamefont
  {Godr{\`e}che}(1999)}]{Drouffe:1999}%
  \BibitemOpen
  \bibfield  {author} {\bibinfo {author} {\bibfnamefont {J.-M.}\ \bibnamefont
  {Drouffe}}\ and\ \bibinfo {author} {\bibfnamefont {C.}~\bibnamefont
  {Godr{\`e}che}},\ }\bibfield  {Phase ordering and persistence in a class of
  stochastic processes interpolating between the Ising and voter models} {\emph
  {\bibinfo {title} {Phase ordering and persistence in a class of stochastic
  processes interpolating between the ising and voter models},\ }}\href
  {http://stacks.iop.org/0305-4470/32/i=2/a=003} {\bibfield  {journal}
  {\bibinfo  {journal} {J. Phys. A}\ }\textbf {\bibinfo {volume} {32}},\
  \bibinfo {pages} {249} (\bibinfo {year} {1999})}\BibitemShut {NoStop}%
\bibitem [{\citenamefont {Mobilia}\ and\ \citenamefont
  {Redner}(2003)}]{Mobilia:2003a}%
  \BibitemOpen
  \bibfield  {author} {\bibinfo {author} {\bibfnamefont {M.}~\bibnamefont
  {Mobilia}}\ and\ \bibinfo {author} {\bibfnamefont {S.}~\bibnamefont
  {Redner}},\ }\bibfield  {Majority versus minority dynamics: Phase transition
  in an interacting two-state spin system} {\emph {\bibinfo {title} {Majority
  versus minority dynamics: Phase transition in an interacting two-state spin
  system},\ }}\href {http://link.aps.org/doi/10.1103/PhysRevE.68.046106}
  {\bibfield  {journal} {\bibinfo  {journal} {Phys. Rev. E}\ }\textbf {\bibinfo
  {volume} {68}},\ \bibinfo {pages} {046106} (\bibinfo {year}
  {2003})}\BibitemShut {NoStop}%
\bibitem [{\citenamefont {{Ugander}}\ \emph {et~al.}(2011)\citenamefont
  {{Ugander}}, \citenamefont {{Karrer}}, \citenamefont {{Backstrom}},\ and\
  \citenamefont {{Marlow}}}]{Ugander:2011}%
  \BibitemOpen
  \bibfield  {author} {\bibinfo {author} {\bibfnamefont {J.}~\bibnamefont
  {{Ugander}}}, \bibinfo {author} {\bibfnamefont {B.}~\bibnamefont {{Karrer}}},
  \bibinfo {author} {\bibfnamefont {L.}~\bibnamefont {{Backstrom}}}, \ and\
  \bibinfo {author} {\bibfnamefont {C.}~\bibnamefont {{Marlow}}},\ }\bibfield
  {{The Anatomy of the Facebook Social Graph}} {\emph {\bibinfo {title} {{The
  Anatomy of the Facebook Social Graph}},\ }}\href@noop {} {\bibfield
  {journal} {\bibinfo  {journal} {{arXiv: 1111.4503}}\ } (\bibinfo {year}
  {2011})}\BibitemShut {NoStop}%
\bibitem [{\citenamefont {Glauber}(1963)}]{Glauber:1963}%
  \BibitemOpen
  \bibfield  {author} {\bibinfo {author} {\bibfnamefont {R.~J.}\ \bibnamefont
  {Glauber}},\ }\bibfield  {Time‐Dependent Statistics of the Ising Model}
  {\emph {\bibinfo {title} {Time‐dependent statistics of the ising model},\
  }}\href
  {http://scitation.aip.org/content/aip/journal/jmp/4/2/10.1063/1.1703954}
  {\bibfield  {journal} {\bibinfo  {journal} {J. Math. Phys.}\
  }\textbf {\bibinfo {volume} {4}},\ \bibinfo {pages} {294} (\bibinfo {year}
  {1963})}\BibitemShut {NoStop}%
\bibitem [{\citenamefont {Fern\'andez-Gracia}\ \emph
  {et~al.}(2014)\citenamefont {Fern\'andez-Gracia}, \citenamefont {Suchecki},
  \citenamefont {Ramasco}, \citenamefont {San~Miguel},\ and\ \citenamefont
  {Egu\'{\i}luz}}]{Fernandez-Gracia:2014}%
  \BibitemOpen
  \bibfield  {author} {\bibinfo {author} {\bibfnamefont {J.}~\bibnamefont
  {Fern\'andez-Gracia}}, \bibinfo {author} {\bibfnamefont {K.}~\bibnamefont
  {Suchecki}}, \bibinfo {author} {\bibfnamefont {J.~J.}\ \bibnamefont
  {Ramasco}}, \bibinfo {author} {\bibfnamefont {M.}~\bibnamefont {San~Miguel}},
  \ and\ \bibinfo {author} {\bibfnamefont {V.~M.}\ \bibnamefont
  {Egu\'{\i}luz}},\ }\bibfield  {Is the Voter Model a Model for Voters?} {\emph
  {\bibinfo {title} {Is the voter model a model for voters?}\ }}\href
  {http://link.aps.org/doi/10.1103/PhysRevLett.112.158701} {\bibfield
  {journal} {\bibinfo  {journal} {Phys. Rev. Lett.}\ }\textbf {\bibinfo
  {volume} {112}},\ \bibinfo {pages} {158701} (\bibinfo {year}
  {2014})}\BibitemShut {NoStop}%
\bibitem [{\citenamefont {DellaVigna}\ and\ \citenamefont
  {Kaplan}(2007)}]{DellaVigna:2007}%
  \BibitemOpen
  \bibfield  {author} {\bibinfo {author} {\bibfnamefont {S.}~\bibnamefont
  {DellaVigna}}\ and\ \bibinfo {author} {\bibfnamefont {E.}~\bibnamefont
  {Kaplan}},\ }\bibfield  {The Fox News Effect: Media Bias and Voting} {\emph
  {\bibinfo {title} {The fox news effect: Media bias and voting},\ }}\href
  {http://qje.oxfordjournals.org/content/122/3/1187.abstract} {\bibfield
  {journal} {\bibinfo  {journal} {The Quarterly Journal of Economics}\ }\textbf
  {\bibinfo {volume} {122}},\ \bibinfo {pages} {1187} (\bibinfo {year}
  {2007})}\BibitemShut {NoStop}%
\end{thebibliography}%

\end{document}